\documentclass[12pt]{iopart}

 \usepackage[utf8]{inputenc}
 \usepackage[T1]{fontenc}
\usepackage{color}

\usepackage{bm}
\usepackage{amssymb}
\usepackage{graphicx}
\usepackage{array}
\usepackage{tabulary}

\usepackage{etoolbox}
\apptocmd{\sloppy}{\hbadness 10000\relax}{}{}

\newcolumntype{K}[1]{>{\centering\arraybackslash}p{#1}}

\begin{document}

\title{Density functional theory for nuclear fission -- a proposal}

\author{J. Dobaczewski}
\address{Department of Physics, University of York, Heslington,
York YO10 5DD, United Kingdom}
\address{Institute of Theoretical Physics, Faculty of Physics,
University of Warsaw, ul. Pasteura 5, PL-02-093 Warsaw, Poland}
\address{Helsinki Institute of Physics, P.O. Box 64, FI-00014 University of Helsinki, Finland}

\begin{abstract}
The fission process is a fascinating phenomenon in which the atomic
nucleus, a compact self-bound mesoscopic system, undergoes a
spontaneous or induced quantum transition into two or more fragments.
A predictive, accurate and precise description of nuclear fission,
rooted in a fundamental quantum many-body theory, is one of the
biggest challenges in science. Current approaches assume adiabatic
motion of the system with internal degrees of freedom at thermal
equilibrium. With parameters adjusted to data, such modelling works
well in describing fission lifetimes, fragment mass distributions, or
their total kinetic energies. However, are the assumptions valid? For
the fission occurring at higher energies and/or shorter times, the
process is bound to be non-adiabatic and/or non-thermal.  The vision
of this project is to go beyond these approximations, and to obtain a
unified description of nuclear fission at varying excitation
energies. The key elements of this project are the use of nuclear
density functional theory with novel, nonlocal density functionals
and innovative high-performance computing techniques. Altogether, the
project aims at better understanding of nuclear fission, where slow,
collective, and semi-classical effects are intertwined with fast,
microscopic, quantum evolution.
\end{abstract}

\maketitle

\section{Introduction and background}

The vision of this proposal is to bring into the physics of nuclear
fission the most advanced theoretical ideas and computation. Since
the discovery of fission almost eighty years ago [Hah38,Hah39,Mei39],
a wealth of experimental data has been accumulated. This has been
accompanied by the development of an efficient phenomenology and
microscopy of spontaneous and induced fission
[Wag91,Sim12,Sch16,And18]. However, almost all of these studies rely
on assuming the adiabaticity and/or thermalisation of fission. Is the
energy sufficiently low and time sufficiently long for these
assumptions to hold?  This project has ambition to implement
theoretical modelling of fission that will deliver definite answers
to these challenges.

From the outside, fission looks like a simple process where a single
drop of matter splits into two or more smaller drops. However, this
is very misleading: a huge conceptual gap exists between the
splitting of liquid drops and nuclear fission. Briefly, during the
fission process, one composite quantum system splits into two or more
composite quantum systems, and all properties of such a process
crucially depend on quantum physics, which is not the case for the
classical liquid drop. Here, nucleons move in correlated quantum
orbitals that evolve into correlated quantum orbitals within the
fission fragments. Altogether, in fission we find all the beauty and
difficulty of a mesoscopic system. It happens in the border region
between classical and quantal, large and small, and slow and fast
phenomena. This is why it is so challenging and consequently provides
an important subject of fundamental science research.

Currently, the methodology used for describing induced fission
[Sch13,And18] at varying excitation energies is in a very rudimentary
state. The standard framework, dating all the way back to the
pioneering work of Bohr\&Wheeler in 1939 [Boh39], is that of a hot
thermalized compound nucleus, which is created after resonant neutron
capture [Sch15,Sch16]. However, applying this concept to the other
mechanisms of creating pre-fission states [And18] is not really the
right way to proceed. Indeed, after the beta-decay of a precursor
system [And13], photon absorption, Coulomb excitation by a passing
charge, or particle transfer, the nucleus ends up in a fairly well
determined intermediate "doorway" state, which then fissions.
Depending on the excitation energy and fission time scale, such an
intermediate state may or may not have enough time to thermalize, and
then the very concept of a compound nucleus becomes highly
questionable.

The main idea is thus to build the doorway states explicitly, by
employing the high-energy vibrational limit of the time dependent
density functional theory (DFT), and then to follow the fission with
coupling to such vibrations included. Indeed, known excitation
operators acting on the pre-fission system can very efficiently model
these states. Then, one should follow the time-evolution of such
intermediate states towards fission. Here, the important challenge
will be to describe the quantum fluctuations that build up
dynamically along the pathway to fission, including those during the
tunnelling process. Both steps will require state-of-the-art computer
technology. Indeed, it is the formidable progress in high performance
computing, together with the recent paradigm shifts in nuclear DFT,
which make this project feasible today.

There are certainly regimes of fission where this phenomenon is
likely to be an adiabatic and thermal process, e.g., after thermal
neutrons are absorbed through long-living resonances [Boh39].
However, when other probes induce fission, it is often probably not.
Indeed, one can expect an entire spectrum of conditions, where
assumptions of adiabaticity and thermalisation are fulfilled or not
to a varying degree. The novel strategy proposed in this project is
to build approaches not relying on these assumptions -- this is
certainly a prerequisite of being able to tell the difference.

\section{The work plan}

Specific objectives of the project are delineated below
within three work packages (WP). They specify how the project will tackle
challenges, formulate the focus and scope of the project, and define
its coherence.

\begin{itemize}

\item[WP1] \underline{Fission with novel functionals in two-centre 3D basis}

\begin{itemize}
\item[(A1)]    For the novel, nonlocal density functionals, solve static superfluid DFT equations for all shapes of fissioning nuclei using a two-centre 3D harmonic-oscillator basis with arbitrary relative distances and orientations;

\item[(A2)]    Implement for these solutions the determination of multi-dimensional fission paths by the minimisation of the collective action within a full, unabridged adiabatic approach;

\item[(A3)]    Implement in these solutions stochastic evolution based on the Langevin method.
\end{itemize}

\item[WP2] \underline{Beyond adiabatic fission}

\begin{itemize}
\item[(B1)]    Using advanced iterative methods, solve linear-response equations for arbitrary multipolarity in all isospin and particle-exchange channels;

\item[(B2)]    Devise and implement novel technology based upon adaptive time-dependent bases to solve the time-dependent DFT (TDDFT) equations, without assuming adiabaticity or linearization;

\item[(B3)]    Solve the problem of adiabatic/diabatic evolution in even and odd fermion systems.
\end{itemize}

\item[WP3] \underline{Beyond thermal fission}

\begin{itemize}
\item[(C1)]    Devise and solve the TDDFT evolution with explicit coupling to two-quasiparticle correlations at varying excitation energies.

\item[(C2)]    Link the TDDFT evolution to instanton solutions in classically forbidden regions of the phase space.
\end{itemize}
\end{itemize}

The work packages logically divide the project into three interwoven
strains with increasing difficulty/high-risk content. They also
divide the objectives roughly into three classes of: feasible
(A1-A3), likely feasible (B1-B3), and maybe feasible (C1-C2) tasks.
For the purpose of organizing the work, the aforementioned objectives
are presented in terms of specific theoretical developments. These
developments will then be followed by concrete steps in code
development and computing.

Below we proceed with a detailed discussion of what the current state
of the art is and how the objectives of this project will go well
beyond the state of the art. We also show why realizing them
constitutes a logical sequence of steps to address the important
challenge, and how these steps will be realized. Then we proceed by
showing high-impact outcomes of the objectives, and we summarize this
section by sketching the coherent big picture of the project.

\subsection{Objective A1: Two-centre 3D code}

The goal is to produce a new DFT solver, which will become the
backbone of the entire project. In nuclear physics, the purpose of a
DFT solver is to deliver self-consistent solutions for a given class
of density functionals, conserved or broken symmetries, numerical
precision, and representation of the Kohn-Sham orbitals [Ben03,Sch19].
Numerous state-of-the-art nuclear superfluid DFT solvers exist
[Ber85,Egi97,Sto05,Ben05,Car10a,Sto13,Nik14,Mar14,Rys15,Jin17,Sch17,Nav17,Dob19],
and some of them have been published as Open Source codes. The type
of solver to be used should be adapted to the physics problem at
hand. For fission, a two-centre solver appears to be an absolute
necessity [Sch16]. The only existing (unpublished, not Open Source)
solver of this type capable of treating a nonlocal (Gogny)
functional, is restricted to the co-axial symmetry of the fission
fragments [Ber80,Gou05,Dub08]. This project will abundantly leverage
the existing technology of the one-centre 3D code HFODD [Sch17,Dob19]
and the new code will be published under Open Source licence.

Objective A1 will develop a two-centre 3D solver with the two
fragments arbitrarily spaced, deformed, and oriented in space, and
for the novel, nonlocal density functionals [Ben17a]. Here and below
pairing correlations will always be included. As is the case in
HFODD, the new solver will be automatically capable of calculating
transition densities and thus the non-diagonal matrix elements of
energy. Here again the fact that the novel, nonlocal density
functionals are based on density-independent generators will make an
essential difference [Dob07]. The non-diagonal matrix elements can
then be easily fed into the symmetry-restoration modules of HFODD, or
into existing codes solving the time-dependent generator coordinate
method in the adiabatic limit of large amplitude collective motion
[Gou04,Reg16,Reg18]. The 3D feature of the new solver will be
essential for studying relative-angular-momentum generation on the
path to scission [Bon07,Ber19a]. The main challenge in Objective A1
is to develop efficient numerical algorithms capable of delivering
the required precision of solutions within a manageable time and
memory, so that the solver can efficiently be used in large-scale
calculations, and then extended to a time-dependent version in
Objective B2.

Objective A1 will implement methodology based on three jointly used elements:

\begin{enumerate}

\item[(i)]     Localised bases, which describe nuclear densities more
efficiently than space lattices. Indeed, space lattices are very
convenient and flexible, but they must treat large volumes of space
(20250 fm3 in [Bul16]), of which the nucleus occupies only a small
fraction;

\item[(ii)]    Two-centre bases, which are vital for describing
fission [Mar72,Sch16], including the essential details of the neck
that bridges fragments at the scission point. In the novel two-centre
3D code, by changing the distance between the centres as well as
shapes and relative orientations of the bases, this project has a
potential to faithfully render fine details of nuclear densities
along the fission pathway;

\item[(iii)]   Harmonic-oscillator bases, which have very simple
analytical properties [Ber80] and allow for performing major parts of
calculations analytically. Such bases are routinely used in our
existing codes [Sch17,Dob19]. The experience thus gained gives us
high confidence and know-how to implement them in the time-dependent
and two-centre settings. In fact, when using novel nonlocal
functionals with exchange terms treated exactly, such bases are, in
practice, the only available option. \end{enumerate}

\subsection{Objective A2: Fission paths}

The goal is to determine spontaneous and induced fission properties
as described by the novel, nonlocal density functionals and within
the two-centre 3D basis. It will pave the way for Objectives C1-C2,
where the same physics goals will be achieved using proposed
high-risk innovative approaches. Here, we will limit our
investigations to determining adiabatic collective paths, with
induced fission treated within the standard state-of-the-art thermal
approximation. This will allow us to benchmark the novel components
of our approach (nonlocal functionals, two-centre 3D solver) against
the most advanced results available in the literature
[Dub08,Pei09,Ich12,Mol12,War12,McD14,Rod14,Sch14,Sch15,Giu14,Zha15,Bar15,Lem15,\newline
Sad16,Pas16,And16,Tan17].
The novelty of this objective will be in: (i) simultaneously treating
all degrees of freedom relevant to fission: elongation, triaxiality,
asymmetry, necking, and pairing; and (ii) using the exact adiabatic
inertia tensor without any simplifying approximations. Especially the
dynamic treatment of pairing correlations will be of primary
importance [Vaq13,Sad14,Giu14,Zha16,Rod18,Ber19b]. At the same time,
we will be able to test the validity conditions of the adiabatic
approximation [Rei78,Rin80].

The use of novel functionals will be an essential aspect of the
approach, as they describe particle-hole and pairing channels on the
same footing and without any density dependence. Massive
large-scale calculations will be performed, providing high-impact
results in the form of systematics of fission properties in different
nuclei and at different excitation energies. This will be the first
calculation that informs novel functionals about their overall
performance at extreme deformations.

Objective A2 will use the methodology already developed in [Sad14]
and implemented in the code HFODD. This will allow us to calculate
fission properties for novel, nonlocal density functionals
systematically, while using the full set of unconstrained space
degrees of freedom and pairing. To determine the complete mass
tensor, an unabridged adiabatic approximation will be implemented.
For that, we will use the technology based on iterative Arnoldi-type
[Toi10] solution of the adiabatic equations, as proposed in [Dob81].

\subsection{Objective A3: Langevin method}

The goal is to determine fission-fragment distributions using
state-of-the-art description of fission paths and dissipation. Within
DFT, a Langevin methodology has recently been developed [Sad16]. The
new aspect of this project will be not only in extending these ideas
to novel, nonlocal density functionals and two-centre 3D basis, but
also in identifying the microscopic mechanisms responsible for energy
dissipation (energy transfer from collective degrees of freedom to
multi-particle multi-hole excitations).

\subsection{Objective B1: Linear response}

The goal is to determine strength functions and two-quasiparticle
correlations for arbitrary nuclear shapes, from ground state to large
deformations along the fission path. Excitations of arbitrary
multipolarity, isospin, and particle transfer will be treated in the
coherent setting of quasiparticle random phase approximation (QRPA)
[Rin80]. A number of state-of-the-art QRPA approaches already exist,
see e.g. [Eng99,Kha02,Fra05,Per08,Ter10,Hin13]; but here we need such
an approach for the novel, nonlocal density functionals and/or for
the two-centre 3D basis required for fission, which both will be the
focus of the present project. Objective B1 will also allow us to
evaluate for novel functionals the ground-state beta-decay rates and
giant-resonance properties for arbitrarily heavy deformed nuclei.
These aspects will constitute important deliverables and key
high-impact results for nuclear physics and astrophysics. They will
also provide the first opportunity to test the performance of the
novel, nonlocal density functionals in describing excited states in
nuclei.

Objective B1 will be based on solving the QRPA equations for the same
novel, nonlocal density functionals as those used to solve the DFT
and TDDFT equations. This will be achieved without any symmetry
restrictions. The principal idea that will make such an advanced
development possible is to use iterative methods again. Up to now,
such approaches were implemented mostly within the so-called
finite-amplitude method [Nak07]. The iterative Arnoldi method, which
has been developed in[Toi10,Ves12], is an analogous efficient route
to take.

\subsection{Objective B2: Time-dependent DFT}

The goal is to find solutions of the TDDFT equations [Mar04] for
fission by employing an innovative approach of expanding the
time-dependent Kohn-Sham orbitals into the time-dependent harmonic
oscillator (HO) basis. This idea will generalise the technology
developed in Objective A1 by making parameters of the two-centre 3D
basis depend on time. The TDDFT equations of motion will then
generate not only the time-dependence of the basis-expansion
coefficients, but also the time evolution of the basis parameters
themselves. This hybrid approach will result in the orbitals that
"carry" their own basis with themselves, which is probably the most
efficient way to solve the problem of time evolution. Though elegant,
this idea has never been attempted in the context of nuclear fission
(see [Kno00] for applications in the molecular physics).

The constructed TDDFT code will become the backbone of the rest of
the project. It will allow for: (i) self-consistent generation of
fission paths by TDDFT and subsequent density-constrained DFT
[Uma06]; (ii) systematic calculations of fission paths on the way to
scission; and (iii) for benchmarking adiabatic paths determined in
Objective A2. When ported to the imaginary-time setting, it will
become a baseline tool to implement instanton evolution in
classically forbidden regions of the phase space, which is essential
for Objective C2. The new code will be published under Open Source
licence.

Objective B2 will implement into the physics of fission the approach
analogous to the time-dependent LCAO method, developed for collisions
of sodium cluster ions with caesium atoms [Kno00]. For fission,
sophisticated algorithms developed in [Bul16] allowed for solving the
time-dependent local-density-approximation (LDA) equations with
pairing in 3D space-lattice coordinates. For one fission path in
240Pu, these algorithms require about 16,000 GPU hours. For nonlocal
functionals, that is, beyond LDA, the analogous 3D space-lattice
algorithms have never been developed. Only a 1D space-lattice has
been implemented in [Has13], with the remaining two dimensions still
treated in the HO basis. Rough estimates indicate that for a nonlocal
functional, the full 3D space-lattice implementation would require
about three orders of magnitude more resources than it does for the
LDA [Dob09], that is, at present it is practically impossible to
realize. Therefore, in this project, we propose to use a
time-dependent two-centre 3D HO basis. This will give us a manageable
extension of the two-centre 3D code that will have been developed in
Objective A2..

\subsection{Objective B3: Adiabatic/diabatic}

The goal is to determine fission properties of odd nuclei.
Experimentally, spontaneous-fission lifetimes of odd nuclei are about
five orders of magnitude longer than those of their even-even
neighbours [Hof89]. Such a hindrance is usually attributed to the
so-called specialization energy [Ran73,Kra12,Hes17], by which the
conservation of quantum numbers of the odd nucleon makes the fission
barriers higher. This interpretation must certainly be reconsidered.
The assumption that the odd nucleon follows a diabatic path across
all other orbitals that appear at low energy along the fission path,
is unsustainable. Neither has it so far been backed up by any
realistic calculations. To properly address the problem, the least
one can do is to include the Landau-Zener coupling between orbitals,
see e.g. [Mir08], which unavoidably leads to breaking of symmetries.
Within adiabatic DFT or TDDFT, very little has been done to pursue
such ideas [Sch16]. For example, a very recent state-of-the-art
calculation [Koh17] still involves conservation of axial symmetry.

In this project, we propose a change of paradigm for the
description of fission in odd nuclei. The central hypothesis is that
their longer lifetimes are not related exlusively to increased barrier heights,
but to increased inertia caused by the level crossings and mixings,
cf.\ [Ber94,Bul10]. The infrastructure built in Objectives A1, A2 and
B2, where breaking of all symmetries will have been incorporated,
would allow us to prove this idea rigorously. If this works, we will
be able to perform extensive computations in many odd nuclei.
Needless to say, we will also venture a similar approach to the
fission of odd-odd nuclei to obtain some proof-of-principle results.

Objective B3 will start by addressing the crucial issue of how to
define and implement adiabatic approximation and collective inertia
in odd nuclei. Indeed, within the standard formulation of the
adiabatic approximation, the collective motion is treated through
small time-odd corrections to quasistationary time-even density
matrices [Rin80]. However, in odd nuclei, even stationary density
matrices involve appreciable time-odd components, so the very notion
of adiabaticity, be it still valid or not, needs to be re-defined. We
are confident that such a revamping of adiabatic theory is doable
within this project thanks to his extensive expertise in studies of
the manifestation of time-reversal symmetry breaking in nuclei
[Dob95,Ber09,Sch10].

Thereafter, we will perform the TDDFT calculations in odd nuclei,
where the time-odd density matrices will consistently include
components related both to the presence of an odd fermion and time
dependence itself. This will inform us on how TDDFT should be reduced
to the adiabatic limit in odd nuclei. Finally, we will implement in
the DFT and TDDFT solvers an explicit mixing of states corresponding
to different blocked orbitals. This would follow the strategy of the
no-core configuration interaction, recently proposed in nuclei
[Sat16], which in molecular physics is known as Multi-Configuration
(Time-Dependent) Hartree-Fock method [Fro97,Zan04]. One can foresee
that such approaches can be quite heavy numerically; nevertheless,
they will be an invaluable source if information on how the idea of
non-diabatic Landau-Zaner crossings works in a realistic setting.

\subsection{Objective C1: Correlations}

The goal is to describe nuclear fission without thermal approximation
and thus to obtain a consistent approach to induced fission at
varying excitation energies. This will constitute a high-risk,
all-hands-on-deck part of the project, built upon the solid
foundations laid down by the successful completion of the previous
objectives. The principal break-through idea of how to approach the
problem is the following: The main sector of important correlations
is given by a coupling to two-quasiparticle excitations, see e.g.
[Taj93,Ber11]. These, in turn, are very effectively captured within
the QRPA approach. Moreover, a natural picture of the QRPA is that of
a small-amplitude approximation to TDDFT [Rin80]. Therefore, we will
execute the time evolution along the fission path in such a way that
the small-amplitude vibrations become a part of this evolution.

Specifically, let us assume that we are first interested in the
process of induced fission in the classically allowed region of
energies above the fission barrier [And18] (otherwise see Objective
C2). For such a process, the initial doorway intermediate state can
be created by, e.g., the beta decay [And13] of the parent system,
photon absorption, Coulomb excitation, or particle transfer. In all
these cases, the transition operator corresponds to a well-defined
two-quasiparticle excitation acting on the ground state of the
fissioning nucleus, and it can be very well modelled within QRPA
(Objective B1). Having at our disposal such a time-dependent mode, we
can inject it into the initial condition of TDDFT and follow it on
the way to fission. We can then expect that there will appear a
coupling between the slow collective motion and rapid QRPA modes. In
this way, a part of the collective energy will explicitly dissipate
into the rapid motion, which usually would be interpreted as a
thermalisation or dissipation of the collective energy. Here, the
same physical effect will be rendered through an explicit time
evolution. In addition, we may also observe an inverse feedback of
the rapid motion into the collective evolution, which may speed up
and/or facilitate fission. Another goal of this study will be to
establish links with the ideas based on the overdamped collective
motion [Bul18], which turn out to be fundamental for the description
of the fission towards the scission point.

Recent breakthrough studies showed that the TDDFT calculations
provide meaningful results when used to describe the latter stage of
the fission process [Sim14,Sca15a]. The present project will use
initial condition that correspond to a compact excited state, rather
than to a system outside the fission barrier. Nevertheless, these
recent works indicate that we can expect to be able to deliver
meaningful results of induced fission at different energies.

Objective C1 will rely on the linear-response (Objective B1) and
TDDFT (Objective B2) technology developed previously within this
project. The risks of not being able to reach these two earlier
objectives are limited, and can be further mitigated by a reversal
towards standard, less innovative approaches. In the work towards
reaching Objective C1, there can appear technical difficulties in
obtaining precise solutions for fast and slow motions of the system
coupled together. A complementary study will be to apply the
configuration-interaction approach [Sat16], already discussed in
Objective B3, to couple a few specific QRPA phonons along the fission
path explicitly. This idea seems to be rather straightforward to
implement, in spite of the significant numerical effort required.
Unlike the direct approach based on the TDDFT, this would allow us to
identify specific relevant modes along the fission path. Indeed, the
TDDFT approach will automatically incorporate all modes, and it may
be sometimes difficult to identify the most important ones.

Another, simpler way to implement correlations would be to stick to
the adiabatic limit and use a non-thermal QRPA phonon as a seed
distribution of initial states, from which would stem a distribution
of fragments and total kinetic energies. Altogether, we think that
within the high-risk Objective C1 we will still be able to explore
various ways of incorporating correlations into the fission process.
A more general problem that may appear is that we are trying to
couple fission with truly coherent correlations given by the QRPA
states along the path. This may raise the question of whether the
novel, nonlocal density functionals will be appropriate for that. If
this aspect turned out to be a problem, we would have to go back to
the drawing board of redesigning some properties of these
functionals.

\subsection{Objective C2: Instantons}

The goal is to determine fission lifetimes at energies below the
fission barrier beyond the standard semi-classical approximation, cf.
[Sca15b]. The theory of mean-field time evolution of many-body
systems in classically inaccessible regions of the phase space has
been laid down by Skalski [Ska08]. In this
project, we will attempt implemnting it. We are convinced that the
time is ripe to attack this problem, and to overcome a great
asymmetry between the sophistication of approaches available above
the barrier, and crudity of those below. This will constitute a great
challenge, with a substantial high-risk uncertainty factor. However,
the gain of having obtained the first microscopic estimate of
penetration probability of the multi-dimensional barrier in a
realistic setting is a great argument to go for it.

In fact, the advanced TDDFT solver devised in Objective B2 will give
us a reasonably high chance to succeed. In particular, it will allow
us to attempt implementations of the variational search for instanton
solutions [Ska08]. This project has a decisive advantage over any
hopes of solving the instanton equations before. Indeed, the
instanton solutions require knowing the non-diagonal matrix elements
corresponding to the given functional [Ska08]. The novel, nonlocal
density functionals, which we use here, are density-independent and
thus offer such a possibility.

We will generate the initial conditions for instanton evolution from
the QRPA phonons at a given excitation energy. As a result, the
instanton solution will be coupled to the TDDFT wave packet
(Objective C1) arriving at the inner side of the barrier. Similarly,
at the far end of the barrier, the instanton solution will feed into
the initial states of the outgoing TDDFT+QRPA wave packet. In this
way, we will go beyond the state-of-the-art picture of the TDDFT
pathways to fission, where only a deformation-induced or
boost-induced fission (between saddle and scission) was identified
[God15,God16].

Objective C2 will primarily use the technology of the TDDFT solver
developed in Objective B2 and ported to the imaginary-time setting.
However, before going into numerical implementations, the work plan
will involve a long study period of how one can really approach the
problem. When studying the sub-barrier time evolution, we must begin
by defining what the barrier really is. Indeed, the barrier, or the
potential energy surface (PES), is a concept that clearly pertains to
the adiabatic approximation. The first question to address is thus:
what is the relation between the PES and, e.g., a fraction of the
total energy related to the time-even part of the TDDFT density
matrix. The same question should also be asked for the instanton
evolution: how to relate the "negative" kinetic energy below the
adiabatic barrier to properties of the instanton in the non-adiabatic
limit.

\section{Summary}

The overarching research question addressed in this project is: How
to describe, understand and predict the large amplitude dynamics of
mesoscopic quantum systems. This question will be answered by
building novel theoretical approaches to study the fission process in
nuclei. The research proposed here is curiosity-driven and addresses
important challenging questions in many-body physics that has
potential for providing a new level of understanding of quantum
time-dependent phenomena.

The principal high-impact outcome of the project will be to answer
important scientific questions related to the nature of nuclear
fission: What is the true microscopic nature of the fission process?
To what extent is the fission process adiabatic? At which energies
are the states of the fissioning nucleus thermalized? These answers
will have a tremendous impact on our understanding of fission and on
further investigations of this process. They also have potential to
influence the understanding of dynamics of other mesoscopic systems
like molecules and processes like chemical reactions.

An overarching outcome of the project will be in the construction of
a coherent calculation infrastructure to address numerous static and
dynamic phenomena in nuclei. Contemporary theoretical nuclear physics
critically depends on high-performance computing. The range and scope
of what can be achieved is most often dictated by the efficiency of
implemented algorithms and available computer power. This project
will go well beyond the state of the art in devising innovative
algorithms to solve nuclear DFT and TDDFT problems, with a specific
focus on the phenomenon of nuclear fission. All computer codes built
within the project will be by default published under Open Source
licence, and thus made available to researchers at large for further
applications and development. This outcome of the project will have a
dramatically high-impact on the entire domain and provide a major
boost for further studies within nuclear DFT.

There is one direction of research that has been voluntarily left out
of this project, but for which the project will be a perfect
springboard. Namely, based on Objectives C1 and C2, future
investigations should attack the problem of microscopic treatment of
dissipative dynamics, see e.g. [Rei86,Abe96,
Sur14,Sla15,Lac16,Mir16,Vin17,Bul18]. This project will approach the
problem of dissipation of collective energy into non-collective
excitations by investigating in Objective C1 the explicit coupling
between these modes. However, the general treatment of collisional
correlations would, unfortunately, go beyond the time frame and
manpower of the project.

The project will be a definite step towards a fully quantum
description of the time-evolution of mesoscopic systems. Once such a
step is completed, several other fascinating research directions
open. For example, one could then look into the problem of quantum
interference between adjacent collective pathways to fission: Are
such effects important? Can multi-dimensional barrier penetration be
really approximated by a single collective path? Maybe one should
consider a collective tube of interfering paths around the collective
pathway to properly determine the fission lifetime, or perhaps
competing and interfering pathways that proceed through distinct
regions of the phase space should be taken into account?

Another virgin territory is the research on the quantum entanglement
of the fission fragments. There is little doubt that they are
entangled, both in terms of their particle-number composition,
relative angular momenta, or pairing gauge phases [Bul17]. The real
question is whether these effects are important for observations down
the line. Can they be observed at all? When and how decoherence of
this entanglement happens? Fission fragments may represent a unique
opportunity to look into the quantum entanglement of mesoscopic
systems, that is, they can be as close as it possibly gets to the
Schr{\"o}dinger cat.

\bigskip

This work was partially supported by the STFC Grants No.~ST/M006433/1
and No.~ST/P003885/1, and by the Polish National Science Centre under
Contract No.~2018/31/B/ST2/02220.

\bigskip

\section*{References}

\begin{itemize}
\item[\mbox{[Abe96]}] Y. Abe, S. Ayik, P.-G. Reinhard, and E. Suraud, Phys. Rep. 275. 49 (1996).
\item[\mbox{[Abg14]}] S.E. Agbemava, A.V. Afanasjev, D. Ray, and P. Ring, Phys. Rev. C89, 054320 (2014).
\item[\mbox{[Afa13]}] A.V. Afanasjev, S.E. Agbemava, D. Ray, and P.Ring, Phys. Lett. B726, 680 (2013).
\item[\mbox{[And13]}] A.N. Andreyev, M. Huyse, and P. Van Duppen, Rev. Mod. Phys. 85, 1541 (2013).
\item[\mbox{[And16]}] A. V. Andreev, G. G. Adamian, and N. V. Antonenko, Phys. Rev. C93, 034620 (2016).
\item[\mbox{[And18]}] A.N. Andreyev, K. Nishio, and K.-H. Schmidt, Rep. Prog. Phys. 81, 016301 (2018).
\item[\mbox{[Bar15]}] A. Baran, M. Kowal, P.-G. Reinhard, L.M. Robledo, A. Staszczak, and M. Warda, Nucl. Phys. A944, 442 (2015).
\item[\mbox{[Ben03]}] M. Bender, P.-H. Heenen, and P.-G. Reinhard, Rev. Mod. Phys. 75, 121 (2003).
\item[\mbox{[Ben05]}] K. Bennaceur and J. Dobaczewski, Comp. Phys. Commun. 168, 96 (2005).
\item[\mbox{[Ben14]}] K. Bennaceur, J. Dobaczewski and F. Raimondi, Eur. Phys. J. Web of Conf. 66, 02031 (2014).
\item[\mbox{[Ben17a]}] K. Bennaceur, A. Idini, J. Dobaczewski, P. Dobaczewski, M. Kortelainen, and F. Raimondi, J. Phys. G: Nucl. Part. Phys. 44, 045106 (2017).
\item[\mbox{[Ben17b]}] K. Bennaceur, J. Dobaczewski, and Y. Gao, Proc. of the Sixth International Conference on Fission and Properties of Neutron-Rich Nuclei, arXiv:1701.08062.
\item[\mbox{[Ber09]}] G. Bertsch, J. Dobaczewski, W. Nazarewicz, and J. Pei, Phys. Rev. A79, 043602 (2009).
\item[\mbox{[Ber11]}] R. Bernard, H. Goutte, D. Gogny, and W. Younes, Phys. Rev. C84, 044308 (2011).
\item[\mbox{[Ber15]}] G.F. Bertsch, W. Loveland, W. Nazarewicz, and P. Talou, J. Phys. G: Nucl. Part. Phys. 42, 077001 (2015).
\item[\mbox{[Ber19a]}]        G.F. Bertsch, arXiv:1901.00928.
\item[\mbox{[Ber19b]}]        R. Bernard, S.A. Giuliani, and L.M. Robledo, Phys. Rev. C99, 064301 (2019).
\item[\mbox{[Ber80]}] J.F. Berger and D. Gogny, Nucl. Phys. A333, 302 (1980).
\item[\mbox{[Ber85]}] J.-F. Berger, Approche microscopique auto-consistante des processus nucl{\'e}aires collectifs de grande amplitude {\`a} basse {\'e}nergie. Application {\`a} la diffusion d'ions lourds et {\`a} la fission, Th{\`e}se, Universit{\'e} Paris-Sud, Centre d'Orsay, 1985
\item[\mbox{[Ber94]}] G.F. Bertsch, Nucl. Phys. A574, 169c (1994).
\item[\mbox{[Bog10]}] S.K. Bogner, R.J. Furnstahl, and A. Schwenk, Prog. Part. Nucl. Phys. 65, 94 (2010).
\item[\mbox{[Boh39]}] N. Bohr and J.A. Wheeler, Phys. Rev. 56, 426 (1939).
\item[\mbox{[Bon07]}] L. Bonneau, P. Quentin, and I.N. Mikhailov, Phys. Rev. C75, 064313 (2007).
\item[\mbox{[Bul10]}] A. Bulgac, J. Phys. G: Nucl. Part. Phys. 37, 064006 (2010).
\item[\mbox{[Bul16]}] A. Bulgac, P. Magierski, K.J. Roche, and I. Stetcu, Phys. Rev. Lett. 116, 122504 (2016).
\item[\mbox{[Bul17]}] A. Bulgac, and Shi Jin, Phys. Rev. Lett. 119, 052501 (2017).
\item[\mbox{[Bul18]}] A. Bulgac, S. Jin, K.J. Roche, N. Schunck, and I. Stetcu, arXiv:1806.00694.
\item[\mbox{[Car08]}] B.G. Carlsson, J. Dobaczewski, and M. Kortelainen, Phys. Rev. C78, 044326 (2008).
\item[\mbox{[Car10a]}]        B.G. Carlsson, J. Dobaczewski, J. Toivanen, and P. Vesel{\'y}, Comp. Phys. Commun. 181, 1641 (2010).
\item[\mbox{[Car10b]}]        B.G. Carlsson and J. Dobaczewski, Phys. Rev. Lett. 105, 122501 (2010).
\item[\mbox{[Cas02]}] E. Castellani, Stud. Hist. Philos. Mod. Phys. 33, 251 (2002); arXiv:physics/0101039.
\item[\mbox{[Del10]}] J.-P. Delaroche, M. Girod, J. Libert, H. Goutte, S. Hilaire, S. P{\'e}ru, N. Pillet, and G.F. Bertsch, Phys. Rev. C81, 014303 (2010).
\item[\mbox{[Dob07]}] J. Dobaczewski, M.V. Stoitsov, W. Nazarewicz, and P.-G. Reinhard, Phys. Rev. C76, 054315 (2007).
\item[\mbox{[Dob09]}] J. Dobaczewski, W. Satu{\l}a, B.G. Carlsson, J. Engel, P. Olbratowski, P. Powa{\l}owski, M. Sadziak, J. Sarich, N. Schunck, A. Staszczak, M. Stoitsov, M. Zalewski and H. Zdu{\'n}czuk, Comp. Phys. Comm. 180, 2361 (2009).
\item[\mbox{[Dob12]}] J. Dobaczewski, K. Bennaceur and F. Raimondi, J. Phys. G: Nucl. Part. Phys. G39, 125103 (2012).
\item[\mbox{[Dob14]}] J. Dobaczewski, W. Nazarewicz, P.-G. Reinhard, J. Phys. G: Nucl. Part. Phys. 41, 074001 (2014).
\item[\mbox{[Dob19]}] J. Dobaczewski et al., 2019, to be published.
\item[\mbox{[Dob81]}] J. Dobaczewski and J. Skalski, Nucl. Phys. A369, 123 (1981).
\item[\mbox{[Dob95]}] J. Dobaczewski and J. Dudek, Phys. Rev. C52, 1827 (1995).
\item[\mbox{[Dub08]}] N. Dubray, H. Goutte, and J.-P. Delaroche, Phys. Rev. C77, 014310 (2008).
\item[\mbox{[Egi97]}] J.L. Egido, L.M. Robledo, and R.R. Chasman, Phys. Lett. B393, 13 (1997).
\item[\mbox{[Eng99]}] J. Engel, M. Bender, J. Dobaczewski, W. Nazarewicz, and R. Surman, Phys. Rev. C 60, 014302 (1999).
\item[\mbox{[Erl12a]}]        J. Erler, K. Langanke, H.P. Loens, G. Mart{\'\i}nez-Pinedo, and P.-G. Reinhard, Phys. Rev. C85, 025802 (2012).
\item[\mbox{[Erl12b]}]        J. Erler et al. Nature 486, 509 (2012).
\item[\mbox{[Fen10]}] Th. Fennel, K.-H. Meiwes-Broer, J. Tiggesb{\"a}umker, P.-G. Reinhard, P. M. Dinh, and E. Suraud, Rev. Mod. Phys. 82, 1793 (2010).
\item[\mbox{[Fra05]}] S. Fracasso and G. Col{\'o}, Phys. Rev. C72, 064310 (2005).
\item[\mbox{[Fro97]}] C. Froese Fischer, T. Brage, and P. Jonsson, Computational Atomic Structure: An MCHF Approach (Institute of Physics Publishing, Bristol,1997).
\item[\mbox{[Giu14]}] S.A. Giuliani, L.M. Robledo, and R. Rodr{\'\i}guez-Guzm{\'a}n, Phys. Rev. C90, 054311 (2014)
\item[\mbox{[Giu17]}] S.A. Giuliani, G. Mart{\'\i}nez-Pinedo, and L.M. Robledo, and M.-R. Wu, Acta Phys. Pol. 48, 299 (2017).
\item[\mbox{[Giu18]}] S.A. Giuliani, G. Mart{\'\i}nez-Pinedo, and L.M. Robledo, Phys. Rev. C97, 034323 (2018).
\item[\mbox{[God15]}] P. Goddard, P. Stevenson, and A. Rios, Phys. Rev. C92, 054610 (2015).
\item[\mbox{[God16]}] P. Goddard, P. Stevenson, and A. Rios, Phys. Rev. C93, 014620 (2016).
\item[\mbox{[Gor09a]}]        S. Goriely, S. Hilaire, M. Girod, and S. P{\'e}ru, Phys. Rev. Lett. 102, 242501 (2009).
\item[\mbox{[Gor09b]}]        S. Goriely, N. Chamel, and J.M. Pearson, Phys. Rev. Lett. 102, 152503 (2009).
\item[\mbox{[Gor11]}] S. Goriely, A. Bauswein, and H.-T. Janka, Astr. J. Lett. 738, L32 (2011).
\item[\mbox{[Gor13]}] S. Goriely, J.-L. Sida, J.-F. Lema{\^\i}tre, S. Panebianco, N. Dubray, S. Hilaire, A. Bauswein, and H.-T. Janka, Phys. Rev. Lett. 111, 242502 (2013).
\item[\mbox{[Gor15]}] S. Goriely, Eur. Phys. J. A  51, 22 (2015).
\item[\mbox{[Gou04]}] H. Goutte. P. Casoli, and J.-F. Berger, Nucl. Phys. A734, 217 (2004).
\item[\mbox{[Gou05]}] H. Goutte, J.F. Berger, P. Casoli, and D. Gogny, Phys. Rev. C71, 024316 (2005).
\item[\mbox{[Hah38]}] O. Hahn and F. Strassmann, Naturwissenschaften 26, 755 (1938).
\item[\mbox{[Hah39]}] O. Hahn and F. Strassmann, Naturwissenschaften 27, 11 (1939).
\item[\mbox{[Has13]}] Y. Hashimoto, Phys. Rev. C 88, 034307 (2013).
\item[\mbox{[Hav19]}] T. Haverinen, M. Kortelainen, J. Dobaczewski, and K. Bennaceur, Acta Phys. Pol. B50, 269 (2019).
\item[\mbox{[Hax02]}] W.C. Haxton and T. Luu, Phys. Rev. Lett. 89, 182503 (2002).
\item[\mbox{[Hes17]}] F.P. He{\ss}berger, Eur. Phys. J. A 53, 75 (2017).
\item[\mbox{[Hin13]}] N. Hinohara, M. Kortelainen, and W. Nazarewicz, Phys. Rev. C87, 064309 (2013).
\item[\mbox{[Hof89]}] D.C. Hoffman, Nucl. Phys. A502. 21c (1989).
\item[\mbox{[Hoh64]}] P. Hohenberg and W. Kohn, Phys. Rev. 136, B864 (1964).
\item[\mbox{[Ich12]}] T. Ichikawa, A. Iwamoto, P. M{\"o}ller, and A.J. Sierk, Phys. Rev. C86, 024610 (2012).
\item[\mbox{[Jin17]}] S. Jin, A. Bulgac, K. Roche, and G. Wlaz{\l}owski, Phys. Rev. C95, 044302 (2010).
\item[\mbox{[Jon15]}] R. O. Jones, Rev. Mod. Phys. 87, 897 (2015).
\item[\mbox{[Kha02]}] E. Khan, N. Sandulescu, M. Grasso, and Nguyen Van Giai, Phys. Rev. C66, 024309 (2002).
\item[\mbox{[Kno00]}] O. Knospe, J. Jellinek, U. Saalmann, and R. Schmidt, Phys. Rev. A61, 022715 (2000).
\item[\mbox{[Koh17]}] Meng-Hock Koh, L. Bonneau, P. Quentin, T.V. Nhan Hao, and H. Wagiran, Phys. Rev. C 95, 014315 (2017).
\item[\mbox{[Koh65]}] W. Kohn and L. Sham, Phys. Rev. 140, A1133 (1965).
\item[\mbox{[Kor10]}] M. Kortelainen et al., Phys. Rev. C82, 024313 (2010)
\item[\mbox{[Kor12]}] M. Kortelainen et al., Phys. Rev. C85, 024304 (2012)
\item[\mbox{[Kor14]}] M. Kortelainen et al., Phys. Rev. C89, 054314 (2014)
\item[\mbox{[Kra12]}] H.J. Krappe and K. Pomorski, Theory of Nuclear Fission (Springer-Verlag, Berlin, 2012)
\item[\mbox{[Lac16]}] L. Lacombe, P.-G. Reinhard, P.M. Dinh, and E. Suraud, J. Phys. B: At. Mol. Opt. Phys. 49, 245101 (2016)
\item[\mbox{[Lem15]}] J.-F. Lema{\^\i}tre, S. Panebianco, J.-L. Sida, S. Hilaire, and S. Heinrich, Phys. Rev. C92, 034617 (2015).
\item[\mbox{[Lep97]}] G.P. Lepage, lectures given at the VIII Jorge Andre Swieca Summer School (Brazil, 1997), nucl-th/9706029.
\item[\mbox{[Mar04]}] M.A.L. Marques and E.K.U. Gross, Annu. Rev. Phys. Chem. 55, 427 (2004).
\item[\mbox{[Mar14]}] J.A. Maruhn, P.-G. Reinhard, P.D. Stevenson and A.S. Umar, Comp. Phys. Comm. 185, 2195 (2014).
\item[\mbox{[Mar72]}] J. Maruhn and W. Greiner, Z. Physik 251, 431 (1972).
\item[\mbox{[McD14]}] J.D. McDonnell, W. Nazarewicz, J.A. Sheikh, A. Staszczak, and M. Warda, Phys. Rev. C90, 021302 (2014).
\item[\mbox{[Mei39]}] L. Meitner and O. Frisch, Nature 143, 239 (1939).
\item[\mbox{[Mir08]}] M. Mirea, Phys. Rev. C78, 044618 (2008).
\item[\mbox{[Mir16]}] M, Mirea, J. Phys. G: Nucl. Part. Phys. 43, 105103 (2016).
\item[\mbox{[Mol12]}] P. M{\"o}ller, J. Randrup, and A.J. Sierk, Phys. Rev. C85, 024306 (2012).
\item[\mbox{[Mum16]}] M.R. Mumpower, R. Surman, G.C. McLaughlin, and A. Aprahamian, Prog. Part. Nucl. Phys. 86, 86 (2016).
\item[\mbox{[Nak07]}] T. Nakatsukasa, T. Inakura, and K. Yabana, Phys. Rev. C 76, 024318 (2007).
\item[\mbox{[Nik14]}] T. Nik{\v s}i{\'c}, N. Paar, D. Vretenar, and P. Ring, Comp. Phys. Comm. 185, 1808 (2014).
\item[\mbox{[Pan16]}] I. Panov, Yu. Lutostansky, and F.-K. Thielemann, J. of Phys. Conf. Ser. 665, 012060 (2016).
\item[\mbox{[Pas16]}] H. Pa{\c s}ca, A.V. Andreev, G.G. Adamian, and N.V. Antonenko, Phys. Lett. B760, 800 (2016).
\item[\mbox{[Pei09]}] J.C. Pei, W. Nazarewicz, J.A. Sheikh, and A.K. Kerman, Phys. Rev. Lett. 102, 192501 (2009).
\item[\mbox{[Per08]}] S. P{\'e}ru and H. Goutte, Phys. Rev. C77, 044313 (2008).
\item[\mbox{[Pet12]}] I. Petermann, K. Langanke, G. Mart{\'\i}nez-Pinedo, I.V. Panov, P.-G. Reinhard, and F.-K. Thielemann, Eur. Phys. J. A48, 122 (2012).
\item[\mbox{[Rai11a]}]        F. Raimondi, B.G. Carlsson and J. Dobaczewski, Phys. Rev. C83, 054311 (2011).
\item[\mbox{[Rai11b]}]        F. Raimondi, B.G. Carlsson, J. Dobaczewski and J. Toivanen, Phys. Rev. C84, 064303 (2011).
\item[\mbox{[Rai14]}] F. Raimondi, K. Bennaceur, and J. Dobaczewski, J. Phys. G: Nucl. Part. Phys. 41, 055112 (2014).
\item[\mbox{[Ran73]}] J. Randrup, C.F. Tsang, P. M{\"o}ller, S.G. Nilsson, and S.E. Larsson, Nucl. Phys. A 217, 221 (1973).
\item[\mbox{[Reg16]}] D. Regnier, M. Verri{\`e}re, N. Dubray, and N. Schunck, Comp. Phys. Comm. 200, 350 (2016).
\item[\mbox{[Reg18]}] D. Regnier, M. Verri{\`e}re, N. Dubray, and N. Schunck, Comp. Phys. Comm. 225, 180 (2018).
\item[\mbox{[Rei03]}] P.-G. Reinhard and E. Suraud, Introduction to Cluster Dynamics (Wiley, Oxford, 2003)
\item[\mbox{[Rei78]}] P.G Reinhard  and K Goeke, Nucl. Phys. A312, 121 (1978).
\item[\mbox{[Rei86]}] P.G Reinhard, H Reinhardt, and K Goeke, Ann. Phys.  166, 257 (1986),
\item[\mbox{[Rin80]}] P. Ring and P. Schuck, The Nuclear Many-Body Problem (Springer-Verlag, Berlin, 1980).
\item[\mbox{[Rod14]}] R. Rodr{\'\i}guez-Guzm{\'a}n and L.M. Robledo, Eur. Phys. J. A50, 142 (2014); Phys. Rev. C89, 054310 (2014).
\item[\mbox{[Rod18]}] R. Rodr{\'\i}guez-Guzm{\'a}n and L.M. Robledo, Phys. Rev. C98, 034308 (2018).
\item[\mbox{[Rys15]}] W. Ryssens, V. Hellemans, M. Bender, and P.-H. Heenen, Comp. Phys. Commun. 187, 175 (2015).
\item[\mbox{[Sad14]}] J. Sadhukhan, J. Dobaczewski, W. Nazarewicz, J.A. Sheikh, and A. Baran, Phys. Rev. C90 (2014) 061304(R).
\item[\mbox{[Sad16]}] J. Sadhukhan, W. Nazarewicz, and N. Schunck, Phys. Rev. C93, 011304(R) (2016).
\item[\mbox{[Sat16]}] W. Satu{\l}a, P. B{\c a}czyk, J. Dobaczewski, and M. Konieczka, Phys. Rev. C94, 024306 (2016).
\item[\mbox{[Sca15a]}]        G. Scamps, C. Simenel, and D. Lacroix, Phys. Rev. C92, 011602(R) (2015).
\item[\mbox{[Sca15b]}]        G. Scamps and K. Hagino, Phys. Rev. C91, 044606 (2015).
\item[\mbox{[Sch10]}] N. Schunck, J. Dobaczewski, J. McDonnell, J. Mor{\'e}, W. Nazarewicz, J. Sarich, and M.V. Stoitsov, Phys. Rev. C81, 024316 (2010).
\item[\mbox{[Sch13]}] N. Schunck, J. of Physics: Conf. Ser. 436, 012058 (2013).
\item[\mbox{[Sch14]}] N. Schunck, D. Duke, H. Carr, and A. Knoll, Phys. Rev. C90, 054305 (2014).
\item[\mbox{[Sch15]}] N. Schunck, D. Duke, and H. Carr: Phys. Rev. C91, 034327 (2015).
\item[\mbox{[Sch16]}] N. Schunck and L.M. Robledo, Rep. Prog. Phys. 79, 116301 (2016).
\item[\mbox{[Sch17]}] N. Schunck, J. Dobaczewski, W. Satu{\l}a, P. B{\c a}czyk, J. Dudek, Y. Gao, M. Konieczka, K. Sato, Y. Shi, X.B. Wang, and T.R. Werner, Comp. Phys. Comm. 216, 145 (2017).
\item[\mbox{[Sch19]}] N. Schunck, editor, Energy Density Functional Methods for Atomic Nuclei (IOP Publishing, Bristol, 2019).
\item[\mbox{[Sim12]}] C. Simenel, Eur. Phys. J. A 48, 152 (2012).
\item[\mbox{[Sim14]}] C. Simenel and A.S. Umar, Phys. Rev. C89, 031601(R) (2014).
\item[\mbox{[Ska08]}] J. Skalski, Phys. Rev. C77, 064610 (2008).
\item[\mbox{[Sla15]}] N. Slama, P.-G. Reinhard, and E. Suraud, Ann. Phys. 355, 182 (2015).
\item[\mbox{[Sto03]}] M.V. Stoitsov, J. Dobaczewski, W. Nazarewicz, S. Pittel, and D.J. Dean, Phys. Rev. C68, 054312 (2003).
\item[\mbox{[Sto05]}] M.V. Stoitsov, J. Dobaczewski, W. Nazarewicz, and P. Ring, Comp. Phys. Commun. 167, 43 (2005).
\item[\mbox{[Sto13]}] M.V. Stoitsov, N. Schunck, M. Kortelainen, N. Michel, H. Nam, E. Olsen, J. Sarich, and S. Wilde, Comp. Phys. Comm.184, 1592 (2013).
\item[\mbox{[Sur14]}] E. Suraud and P.-G. Reinhard, New J. Phys. 16, 063066 (2014).
\item[\mbox{[Taj93]}] N. Tajima, H. Flocard, P. Bonche, J. Dobaczewski, and P.-H. Heenen, Nucl. Phys. A551, 409 (1993).
\item[\mbox{[Tan17]}] Y. Tanimura, D. Lacroix, and S. Ayik, Phys. Rev. Lett. 118, 152501 (2017).
\item[\mbox{[Tar14a]}]        D. Tarpanov, J. Dobaczewski, J. Toivanen and B.G. Carlsson, Phys. Rev. Lett. 113, 252501 (2014).
\item[\mbox{[Tar14b]}]        D. Tarpanov, J. Toivanen, J. Dobaczewski, and B.G. Carlsson, Phys. Rev. C89, 014307 (2014).
\item[\mbox{[Ter10]}] J. Terasaki and J. Engel, Phys. Rev. C82, 034326 (2010).
\item[\mbox{[Toi10]}] J. Toivanen, B.G. Carlsson, J. Dobaczewski, K. Mizuyama, R.R. Rodr{\'\i}guez-Guzm{\'a}n, P. Toivanen, and P. Vesel{\'y}, Phys. Rev. C81, 034312 (2010).
\item[\mbox{[Uma06]}] A.S. Umar and V.E. Oberacker, Phys. Rev. C74, 021601(R) (2006).
\item[\mbox{[Vaq13]}] N.L. Vaquero, J.L. Egido and T.R. Rodr{\'\i}guez, Phys. Rev. C88, 064311 (2013).
\item[\mbox{[Ves12]}] P. Vesel{\'y} J. Toivanen, B.G. Carlsson, J. Dobaczewski, N. Michel, and A. Pastore, Phys. Rev. C86, 024303 (2012).
\item[\mbox{[Vin17]}] M. Vincendon, E. Suraud, and P.-G. Reinhard, Eur. Phys. J. D71, 179 (2017).
\item[\mbox{[Wag91]}] C. Wagemans, editor, The Nuclear Fission Process (CRC Press, Boca Raton, 1991).
\item[\mbox{[War12]}] M. Warda, A. Staszczak, and W. Nazarewicz, Phys. Rev. C86, 024601 (2012).
\item[\mbox{[Zan04]}] J. Zanghellini, M. Kitzler, T. Brabec, and A. Scrinzi, J. Phys. B: At. Mol. Opt. Phys. 37, 763 (2004).
\item[\mbox{[Zha15]}] Jie Zhao, Bing-Nan Lu, T. Nik{\v s}i{\'c}, and D. Vretenar, Phys. Rev. C92, 064315 (2015).
\item[\mbox{[Zha16]}] Jie Zhao, Bing-Nan Lu, T. Nik{\v s}i{\'c}, D. Vretenar, and Shan-Gui Zhou, Phys. Rev. C93, 044315 (2016).
\end{itemize}

\bibliographystyle{iopart-num}
\bibliography{/Actual/LaTeX/Latex.all/jacwit37}

\end{document}